\documentstyle[twocolumn,prl,aps,epsfig]{revtex}
\begin{document}
\draft
\twocolumn[\hsize\textwidth\columnwidth\hsize\csname@twocolumnfalse%
\endcsname

\title{High-Field Electrical Transport in Single-Wall Carbon
Nanotubes}

\author{Zhen Yao$^1$, Charles L. Kane$^2$, and Cees Dekker$^1$}

\address{$^1$Department of Applied Sciences and DIMES, Delft University
of Technology,\\ Lorentzweg 1, 2628 CJ Delft, The Netherlands \\
$^2$Department of Physics, University of Pennsylvania,
Philadelphia, PA 19104 }

\date{November 5, 1999} \maketitle

\begin{abstract}
Using low-resistance electrical contacts, we have measured the
intrinsic high-field transport properties of metallic single-wall
carbon nanotubes. Individual nanotubes appear to be able to carry
currents with a density exceeding $10^9~\mbox{A}/\mbox{cm}^2$. As
the bias voltage is increased, the conductance drops dramatically
due to scattering of electrons.  We show that the current-voltage
characteristics can be explained by considering optical or
zone-boundary phonon emission as the dominant scattering mechanism
at high field.
\end{abstract}
 ]

The potential electronic application of single-wall carbon
nanotubes (SWNTs) requires a detailed understanding of their
fundamental electronic properties, which are particularly
intriguing due to their one-dimensional (1D) nature \cite{dekker}.
Metallic SWNTs have two 1D subbands crossing at the Fermi energy.
In the ideal case the resistance is thus predicted to be
$h/4e^{2}$ or 6.5~k$\Omega$.  In early electrical transport
experiments, however, the nanotubes typically formed a tunnel
barrier of high resistance of $\sim$1~M$\Omega$ with the metal
contacts \cite{tans,bockrath}.  Consequently the bias voltage
dropped almost entirely across the contacts, and tunneling
dominated the transport.  A number of interesting phenomena have
been observed in this regime. At low temperatures, Coulomb
blockade effects prevail \cite{tans,bockrath}. At relatively high
temperatures, the transport characteristics appear to be described
by tunneling into the so-called Luttinger liquid --- a unique
correlated electronic state in 1D conductors which is due to
electron interactions \cite{bockrathpostma,yao}.

One of the most important questions that remains to be addressed
is how the electrons traverse the nanotubes, i.e., whether
ballistically or being scattered by impurities or phonons. The
unusual band structure of metallic tubes suggests a suppression of
elastic backscattering of electrons by long-range disorder
\cite{ando}. Long mean free paths for electrons near the Fermi
energy have indeed been inferred from regular Coulomb oscillations
and coherent tunneling at low temperatures \cite{tans,mceuen}.
However, there has been no transport study of electrons with
significant excess energy above the Fermi energy. It is not clear
whether such electrons would experience strong scattering and what
type of scattering mechanism would dominate.

In this Letter we present electrical transport measurements of
individual nanotubes using low-resistance contacts (LRCs). In
contrast to the high-resistance contacts (HRCs), a bias voltage
applied between two LRCs establishes an electric field across the
nanotube which accelerates the electrons, enabling transport
studies of high-energy electrons. We find that individual SWNTs
can sustain a remarkably high current density of more than
$10^9$~$\mbox{A}/\mbox{cm}^2$.  The current seems to saturate at
high electric field.  We discuss possible scattering mechanisms
and suggest that optical or zone-boundary phonon emission by
high-energy electrons can explain the observed behavior. An
analytic theory based on the Boltzmann equation is developed which
includes both elastic scattering and phonon emission. The
numerical calculations reproduce the experimental results
remarkably well.

\begin{figure}[hbt]
\centerline{\epsfig{file=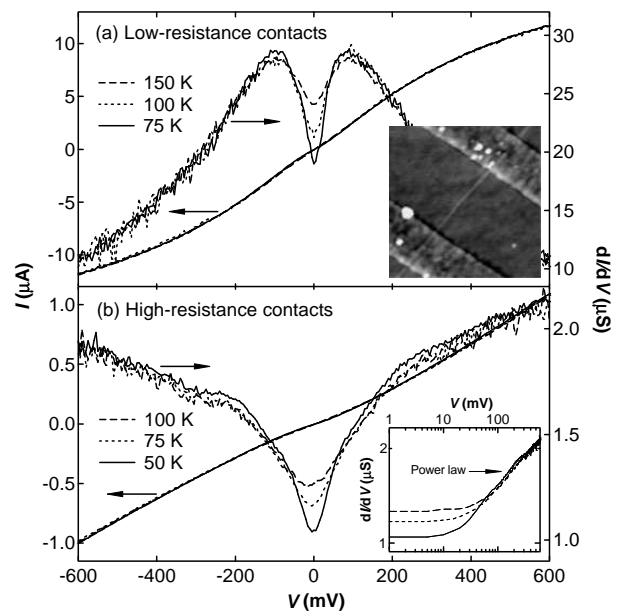, width=8 cm,
clip=}}\vspace{.2cm} \caption{Typical $I$ and $dI/dV$ vs $V$
obtained using (a) low- and (b) high-resistance contacts. The
inset to (a) shows a $0.86 \times 0.86~\mu$m$^2$ AFM height image
of a typical LRC sample, where the $z$ range is 10~nm. The inset
to (b) plots $dI/dV$ vs $V$ on a double-log scale for the HRC
sample.}
\end{figure}

The inset to Fig.~1(a) shows an atomic force microscope (AFM)
image of our typical LRC sample.  The 20~nm thick, 250~nm wide
Ti/Au electrodes are embedded in thermally-grown SiO$_{2}$ with a
height difference of less than 1~nm which minimizes the
deformation of the nanotubes near the electrodes.  This is
achieved by electron-beam lithography and anisotropic reactive-ion
etching of SiO$_{2}$ using a single layer of PMMA as both
electron-beam resist and etching mask, followed by metal
evaporation and lift-off.  The electrodes are cleaned thoroughly
in fuming nitric acid.  The nanotubes are then deposited on top of
the electrodes from a suspension of SWNTs ultrosonically dispersed
in dichloroethane.  We find that brief annealing of the electrodes
at 180$^\circ$C improves the reproducibility of the contact
resistance. Only nanotubes with apparent height of $\sim$1~nm
under AFM are chosen for transport measurements, which are
presumably individual SWNTs.  Metallic nanotubes are selected
based on the absence of gate effect on transport at high
temperatures \cite{tubefet}.  This procedure yields a typical
two-terminal resistance of individual metallic tubes of less than
100~k$\Omega$ (the lowest is 17~k$\Omega$) at room temperature, as
compared to the $\sim$~M$\Omega$ resistance using Pt as the
contact material in previous experiments \cite{tans}. Similar
reduction in contact resistance has also been achieved in a
different contact geometry \cite{soh}. The exact mechanism for the
low contact resistance is unclear. However, clean flat
single-crystalline gold facets may increase the coupling by
increasing the effective contact length over which a small
tube-electrode separation is realized \cite{tersoff}.

Figure~1 shows the typical two-terminal current $I$ and
differential conductance $dI/dV$ vs voltage $V$ obtained using LRC
(Au) and HRC (Pt).  $dI/dV$ is acquired simultaneously using a
standard ac lock-in technique.  The room-temperature zero-bias
resistance of the two samples are 40~k$\Omega$ and 670~k$\Omega$
respectively.  For both samples, the zero-bias conductance $G$
decreases monotonically as the temperature $T$ decreases
\cite{positive}.  The large-bias-voltage dependence of $dI/dV$,
however, is notably different.  For the LRC sample, $dI/dV$
increases with increasing bias, reaching a maximum at
$\sim$100~mV. As the bias increases further, $dI/dV$ drops
dramatically.  In contrast, the HRC sample exhibits a monotonic
increase of $dI/dV$ as a function of voltage up to 1~V. The inset
to Fig.~1(b) plots $dI/dV$ vs $V$ on a double-logarithmic scale
for the HRC sample, in which it appears that $dI/dV$ can be fit
with a power-law function for large bias.  Both the temperature
dependence of $G$ and the bias-voltage dependence of $dI/dV$ for
the HRC sample are typical of individual SWNTs and ropes with
similar or lower conductance values \cite{bockrathpostma}, which
are attributed to the suppressed tunneling density of states in a
Luttinger liquid \cite{eggerkane}.  The similar behavior around
zero bias for the LRC sample suggests that it comes from the same
origin. In the remaining of the paper, we focus on the large-bias
behavior for the LRC samples.

We have further extended the $I$-$V$ measurements up to 5~V as
shown in Fig.~2.  Strikingly, the $I$-$V$ curves at large bias
measured at different temperatures between 4~K and room
temperature essentially overlap with each other. The current at
5~V exceeds 20~$\mu$A, which corresponds to a current density of
more than $10^9$~$\mbox{A}/\mbox{cm}^2$ if a spatial extent of the
$\pi$-electron orbital of $\sim$3~\AA\ is used to estimate the
current-carrying cross section. From the shape of the $I$-$V$
curves, it is clear that the trend of decreasing conductance
continues to high bias. Extrapolating the measured $I$-$V$ curves
to higher voltage would lead to a current saturation, i.e., a
vanishing conductance.  Interestingly, the saturation current
seems to be independent of the distance between the electrodes
\cite{saturation}.

\begin{figure}[hbt]
\centerline{\epsfig{file=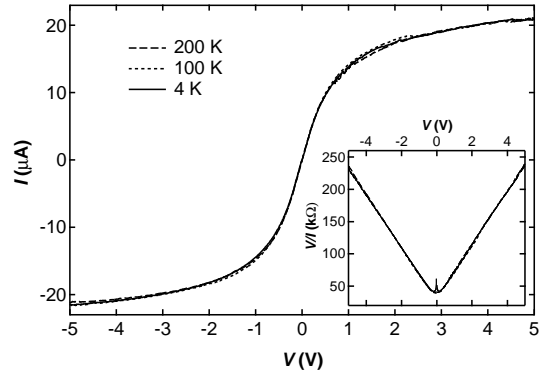, width=7 cm,clip=}}
\vspace{.2cm} \caption{Large-bias $I$-$V$ characteristics at
different temperatures using low-resistance contacts for a sample
with an electrode spacing of 1~$\mu$m.  The inset plots $R \equiv
V/I$ vs $V$.}
\end{figure}

We find that the resistance, $R \equiv V/I$, can be fit remarkably
well by a simple linear function of $V$ over almost the entire
range of applied voltage (see Fig.~2 inset):
\begin{equation}
R = R_0 + V/I_0,
\end{equation}
where $R_0$ and $I_0$ are constants. Only at high voltages near
5~V some samples start showing slight deviation from this linear
behavior. From the slope of the linear part of $R$-$V$ we find
$I_0 \sim 25$~$\mu$A which is approximately the same for all of
the samples we have measured.

At first sight the current saturation might be explained from the
band structure. Current in metallic nanotubes is carried by two
propagating 1D subbands. In the absence of scattering, the
chemical potentials of the right and left moving states will
differ by the applied voltage $eV$.  At low voltages this leads to
an Ohmic response, but when $eV$ exceeds the Fermi energy of the
1D subbands, the left moving states will be completely depleted
and the current will saturate.  The Fermi energy, measured
relative to the nearest band edge, is approximately 2.9~eV.
Experimentally, however, the current starts saturating at a much
lower voltage. An alternative model is thus needed to explain the
saturation.

We expect the measured resistance to be a combination of the
resistance due to the contacts and the resistance due to
backscattering along the length of the nanotube.  The current
saturation is unlikely to arise from an increased contact
resistance at high voltages since the contacts would then behave
like high-resistance tunneling contacts, and one would expect to
see features in the $I$-$V$ associated with tunneling into the 1D
subbands of the nanotube.   The measured $I$-$V$, however, is
featureless.

We thus focus on the effect of backscattering in the nanotube. The
behavior of $R$-$V$ suggests that in addition to a constant
scattering term, which most probably comes from contact scattering
or impurity scattering, there is a dominant scattering mechanism
with a mean free path (mfp) which scales inversely with the
voltage.

Electrons can backscatter off phonons and other electrons.
Electron-electron scattering is appealing at first, since it does
not involve heating the lattice. The only electron-electron
scattering that contributes to resistivity is Umklapp scattering
\cite{balentsfisher} with a scattering rate directly proportional
to the electron temperature $T_{\rm e}$. This gives $V/I \sim
T_{\rm e}$. $T_{\rm e}$ will be determined by how fast the heat
can escape from the tube. If we assume that all the heat produced
is carried by electrons into the leads and that the temperature
along the tube is uniform \cite{umklapp}, we have $IV =
4(\pi^2/3)(k_{\rm B}T_{\rm e})^2/h$, where the left-hand side is
the rate at which heat is produced, and the right-hand side is the
heat current carried by the two 1D channels. Hence we expect that
$I \sim V^{1/3}$, which cannot describe the experiments. We have
verified this $V^{1/3}$ behavior by numerically solving a
Boltzmann equation similar to that discussed below.
Luttinger-liquid effects, which have been ignored in the above
arguments, tend to enhance Umklapp scattering at low
energies\cite{eggerkane}.  This would make the agreement with
experiment even worse.

This suggests that we must consider scattering from phonons. The
coupling will be strongest for phonons which compress and stretch
bonds on the lattice scale.  There are three possible categories
of phonons: (1) twistons or long-wavelength acoustic phonons
\cite{kanemele}, (2) optical phonons which are derived from the
in-plane $E_{2g_2}$ mode of graphite with a frequency of
1580~cm$^{-1}$, and (3) in-plane zone-boundary phonons with
momentum which connects the two Fermi points of graphene.  While
zone-boundary phonons are not directly observable optically,
force-constant models put their frequency in the range
$1000-1500$~cm$^{-1}$ in graphene\cite{fullerene}. Twiston
scattering is most likely not relevant since the scattering rate
is smaller than the optical or zone-boundary phonon scattering by
$\sim T/\Theta_D$, where $T$ is the lattice temperature and
$\Theta_D \sim 2000$~K is the Debye temperature \cite{kanemele}.
It is unlikely that the lattice temperature could be that high.
Moreover, twistons may be pinned by the substrate.

Now we discuss backscattering due to the emission of optical or
zone-boundary phonons.  A related effect has been discussed
previously in the context of semiconductors\cite{mahan}. The key
point is that for an electron with energy $E$ to emit a phonon of
energy $\hbar\Omega$, there must be an available state to scatter
into at energy $E-\hbar\Omega$. In the presence of an electric
field ${\cal E}$, electrons are accelerated, $\hbar \dot{k} = e
{\cal E}$. It is simplest to consider the case in which the
coupling to the phonon is so strong that, once an electron reaches
the threshold for phonon emission, it is immediately
backscattered. As indicated in the schematic in the inset to
Fig.~3, a steady state population is then established in which the
right moving electrons are populated to an energy $\hbar\Omega$
higher than the left moving ones.  The current carried in this
state can be computed from a Landauer type argument to be
\begin{equation}
I_0 = (4e/h)\hbar\Omega.
\end{equation}
If we choose $\hbar\Omega=0.16$~eV (corresponding to 1300
cm$^{-1}$), this leads to a saturation current of 25~$\mu$A, which
is independent of sample length and agrees very well with the
measured saturation current.

In this picture, the mfp for backscattering phonons
$\ell_{\Omega}$ is equal to the distance an electron must travel
to be accelerated to an energy above the phonon energy:
$\ell_{\Omega} = \hbar\Omega/e{\cal E}$. This may be combined with
a constant elastic scattering term via Mathiesson's rule to obtain
an effective mfp, $\ell_{\rm eff}^{-1} = \ell_{\rm e}^{-1} +
\ell_{\Omega}^{-1}$, where $\ell_{\rm e}$ is the elastic
scattering mfp.  The resulting resistance, $R =
(h/4e^2)(L/\ell_{\rm eff})$, then has the empirically observed
form of Eq.~(1) with $R_0 = (h/4e^2)L/\ell_{\rm e}$ and $I_0$
given in Eq.~(2).

To put the above interpretation on a more quantitative basis, we
consider the Boltzmann equation for the distribution functions
$f_{L,R}(E_k,x,t)$ of left and right moving $(L,R)$ electrons.
Details will be provided in a future publication.
\begin{equation}
\left[\partial_t \pm v_F \partial_x \pm v_F  e {\cal
E}\partial_E\right] f_{L,R} = \left[
\partial_t f_{L,R}\right]_{\rm col}.
\end{equation}
Here $v_F$ is the Fermi velocity, and we have chosen to express
the momentum dependence of $f$ in terms of $E_k = \pm \hbar v_F
k$. The left-hand side describes the collisionless evolution of
the electrons in the presence of an electric field ${\cal E}$. For
the collision term on the right, we consider a sum of three terms:
(1) Elastic scattering, $\left[\partial_t f_L\right]_{\rm e} =
(v_F/ \ell_{\rm e})(f_R - f_L)$, where $\ell_{\rm e}$ is the
elastic mfp. (2) Backscattering from phonons, $\left[\partial_t
f_L\right]_{\rm pb} = (v_F /\ell_{\rm pb})\left[(1-f_L)f^+_R -
f_L(1-f^-_R)\right]$. Here $f^\pm$ are evaluated at $E \pm
\hbar\Omega$. $\ell_{\rm pb}$, which depends on the strength of
the electron-phonon coupling, is the distance an electron travels
before backscattering once the phonon emission threshold has been
reached. This should be contrasted with $\ell_{\Omega}$, the
distance required to reach the threshold. We assume that the
phonon temperature is much less than the phonon energy of
$\sim$2000~K, so that the Bose occupation factors can be ignored.
Finally, we consider (3) forward scattering from phonons,
$\left[\partial_t f_L\right]_{\rm pf} = (v_F/ \ell_{\rm
pf})\left[(1-f_L)f^+_L - f_L(1-f^-_L)\right]$.

The effects of the contact resistance may be included as a
boundary condition at the ends of the tube. For instance at the
left contact ($x=0$) we have,
\begin{equation}
f_{R}(E,0)=t_L^2 f_{0}(E-\mu_L)+(1-t_L^2)f_{L}(E,0),
\end{equation}
where $t_L^2$ is the transmission probability for the contact, and
$f_0(E-\mu_L)=(\exp[(E-\mu_L)/k_{\rm B}T]+1)^{-1}$ is the Fermi
distribution function of the left contact with electrochemical
potential $\mu_L$ and temperature $T$.

\begin{figure}[hbt]
\centerline{\epsfig{file=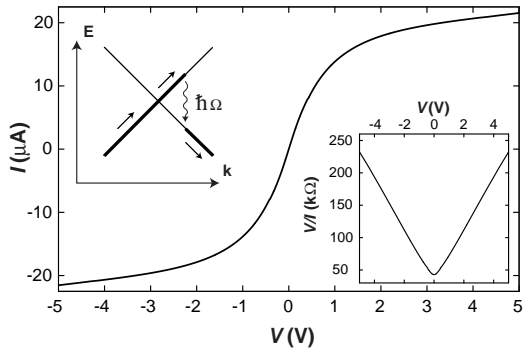, width=7 cm,
clip=}}\vspace{.2cm} \caption{Numerical calculation of $I$-$V$
characteristic by solving Boltzmann tranport equation including
elastic impurity scattering and phonon emission as depicted in the
left inset. See text for the parameters used.}
\end{figure}

We have solved Eqs.~(3,4) numerically to obtain the steady state
distribution function $f_{L,R}(E,x)$ in the presence of an applied
voltage $\mu_L - \mu_R = eV$ as a function of $\hbar\Omega$,
$\ell_{\rm e}$, $\ell_{\rm pb,pf}$, and $t_{L,R}^2$. The current
is then simply given by $I = (4 e/h)\int dE (f_L - f_R)$. Figure~3
shows an example of the numerical calculation of $I$-$V$
characteristic for a sample length of $L=1~\mu$m.  The parameters
used in the plot are $\hbar\Omega = 0.15$~eV, $t_{L,R}^2 =0.5$,
$\ell_{\rm e}=300$~nm, $\ell_{\rm pb} =10$~nm, and $\ell_{\rm pf}
= \infty$. The resemblance to the experiment is remarkable. It is
interesting to note that the current is insensitive to the contact
scattering for $V \gtrsim 0.5$~V. Contact scattering affects only
the low bias resistance, giving rise to the positive curvature in
$R$-$V$ near $V = 0$.

Assuming local thermal equilibrium the Boltzmann equation may be
used to derive hydrodynamic equations which govern the transport
of charge and energy.  These equations may then be solved
analytically and give results which agree well with the
simulations.  They show that: (i) The empirical formula [Eq.~(1)]
is exact in the limit $e V \ll \hbar\Omega L/\ell_{\rm pb}$, which
means that the energy gained by an electron within distance
$\ell_{\rm pb}$ must be much less than the phonon energy, or
equivalently, $\ell_{\rm pb} \ll \ell_\Omega$. (ii) For larger $V$
the simple formula breaks down, and in the limit of very large
$V$, the resistance becomes constant, $R =(h/4e^2)L(\ell_{\rm
e}^{-1} + \ell_{\rm pb}^{-1})$. For the parameters used in Fig.~3,
the crossover voltage is roughly 15~eV. Indeed, there appears a
small negative curvature at 5~V in the $V/I$ vs $V$ plot (inset
Fig.~3), which signals the beginning of the breakdown of the
empirical formula. The curvature would be less pronounced if a
shorter value for $\ell_{\rm pb}$ is used.  We note that 10~nm
seems rather short.  An estimate using a simple deformation
potential model gives $\ell_{\rm pb} \sim 150$~nm for a nearly
armchair nanotube.   More work is needed to have a more accurate
estimate of the electron-phonon coupling strength.

We have assumed that the heat generated in the tube escapes
sufficiently quickly to avoid raising the lattice temperature too
high. A simple estimate of the nanotube's thermal conductivity
indicates that it is unlikely that all of the heat could be
transmitted through the contacts. However, the nanotube is in
intimate contact along its entire length with the substrate, which
may be regarded as a thermal reservoir at $T \leq 300$~K.  It
would clearly be desirable to study further the nature of the
thermal contact between the nanotube and substrate. Measurements
on suspended nanotubes may provide some useful information.

We thank R.E. Smalley and coworkers for providing the
indispensable nanotube materials, M.P.~Anantram, H.~Postma and
S.J.~Tans for discussions, and A.~van den Enden for technical
assistance.  The work at Delft was supported by the FOM and the
work at Pennsylvania by the NSF under grant DMR 96-32598.


\begin{references}
\bibitem{positive} We have also observed an increasing conductance as
temperature decreases for those samples with a two-probe
resistance of less than 25~k$\Omega$ at room temperature, similar
to that found in Ref.~\cite{soh}.
\bibitem{umklapp} The temperature is uniform when the system is
effectively at half-filling across the length of the tube.
\end{references}
\end{document}